\documentclass{mg11}

\newcommand{\be}{\begin{equation}}
\newcommand{\ee}{\end{equation}}
\newcommand{\bea}{\begin{eqnarray}}
\newcommand{\eea}{\end{eqnarray}}

\begin{document}



\title{BIG-RIP, SUDDEN FUTURE, AND OTHER EXOTIC SINGULARITIES IN THE UNIVERSE}

\author{MARIUSZ P. D\c{A}BROWSKI\footnote{Presenting author.} and ADAM BALCERZAK}

\address{Institute of Physics, University of Szczecin,\\
Wielkopolska 15, 70-451 Szczecin, Poland\\
\email{mpdabfz@sus.univ.szczecin.pl}}


\begin{abstract}
We discuss exotic singularities in the evolution of the universe motivated
by the progress of observations in cosmology. Among them there are:
Big-Rip (BR), Sudden Future Singularities (SFS), Generalized
Sudden Future Singularities (GSFS), Finite Density Singularities
(FD), type III, and type IV singularities. We relate some of these
singularities with higher-order characteristics of expansion such
as jerk and snap. We also discuss the behaviour of pointlike
objects and classical strings on the approach to these
singularities.
\end{abstract}

\bodymatter

\section{Introduction}\label{intro}

\noindent
Through many years in the past only the two basic cosmological type of singularities
were known among the isotropic models of the universe. These were Big-Bang and Big-Crunch
appended by a future asymptotic (and non-singular) state of a de-Sitter type. The
appearance of Big-Bang and Big-Crunch was in no way related to any
of the energy conditions violation. The progress in cosmological
observations at the turn of the 21st century \cite{supernovaeold} did not add anything
new to the picture apart from the fact that then it was realized
that these singularities could emerge also in the strong-energy-condition-violation
cases of $\varrho + 3p < 0$. However, a deeper analysis
of the data from supernovae, cosmic microwave background (WMAP) and
large-scale structure (SDSS) \cite{supernovaenew} shows that there
exists other possibilities of the universe evolution which admit new
type of singularities and the problem of the link between energy conditions violation
and the singularity appearance becomes unclear. We will discuss
these new singularities and the problems to relate them with the
possible generalized energy conditions as well as some new
observational characteristics of the expansion of the universe.

\section{Phantom-driven Big-Rip. Phantom duality.}
\noindent
The main motivation to exotic singularities comes from phantom
\cite{caldwell}. Apparently, it emerged that the observational
data does not make any ``borderline'' at $p=-\varrho$ in cosmology
and that the smaller pressure is allowed to dominate current
evolution. Phantom may easily be simulated by a scalar field $\phi$ of negative
kinetic energy which gives the energy-momentum tensor for a
perfect fluid with the energy density $\varrho = -(1/2) \dot{\phi}^2 + V(\phi)$~,
and the pressure $p = -(1/2) \dot{\phi}^2 - V(\phi)$~, so that it
surely violates the null energy condition since $\varrho + p = -\dot{\phi}^2
< 0$.  Phantom is allowed in Brans-Dicke theory in the Einstein frame
(for Brans-Dicke parameter $\omega < -3/2$), in superstring cosmology,
in brane cosmology, in viscous cosmology and many others.
The most striking consequence of phantom is that its
energy density $\varrho$ grows proportionally to the scale factor
$a(t)$. Then, unlike in a more intuitive standard matter case, where the growth of the
energy density corresponds to the decrease of the scale factor,
here, the growth of the energy density accompanies the expansion of the
Universe. This allows a new type of singularity in the universe which is called a Big-Rip.
This singularity appears {\it despite} all the energy conditions
are violated. It is a true singularity in the sense of geodesic incompletness
apart from some range of the possible equations of state for
isotropic geodesics which are complete \cite{lazkoz}. A very
peculiar feature of phantom models against standard models is phantom duality
\cite{phantomcl}. It is a new symmetry of the field equations which allows to map
a large scale factor onto a small one and vice versa due to a
change
\begin{equation}
a(t) \leftrightarrow \frac{1}{a(t)} \hspace{0.5cm} {\rm or } \hspace{0.5cm} w+1 \leftrightarrow
-(w+1)~,
\end{equation}
with a consequence of replacing energy conditions violating matter
onto a non-violating one.

\section{Sudden (and Generalized) Future Singularities, Finite Density singularities, type
III and IV singularities.}
\noindent
Big-Rip leads to violation of all the energy conditions. It
appears that one is able to get some other exotic singularities
which violate the dominant energy condition ($p < \mid \varrho \mid$) only
or even do not violate any energy condition. The former are SFS
and the latter are GSFS. The idea to get them is not to constrain
the set of cosmological field equations by any equation of state
\cite{sudden}, which allows an independent evolution of the energy
density and the pressure. Actually, the energy density depends on at most first derivative
of the scale factor, while the pressure depends on the second
derivative, too. Then, it may happen that at a certain moment of
the evolution only the second derivative of the scale factor is
divergent - this is a Sudden Future Singularity - the energy density remains
finite, while the pressure blows-up to infinity. It was shown that
it is a weak singularity \cite{lazkoz} in the sense of the formal
definitions of singularities known in general relativity. The main
point is that there is no geodesic incompletness at this
singularity and the evolution of an individual pointlike object
can be extended through it. Same refers to Generalized Sudden
Future Singularities. These singularities are temporal (appear at
some fixed time on a hypersurface $t=$ const.), but there exist
also a spatial pressure singularities (may exist somewhere in the universe
nowadays) in cosmology, though in inhomogeneous models \cite{FD}.
It is possible to have inhomogeneous models of the universe which
exhibit both types of singularities. Finally, other exotic types
of singularities are also possible \cite{nojiri}. These are type
III (with finite scale factor and blowing-up the energy density
and pressure) and type IV (with finite scale factor, vanishing the
energy density and pressure, blowing-up the pressure derivative).
It is interesting to know the difference between the evolution of
pointlike objects and extended objects such as fundamental strings
through these various exotic singularities \cite{strings}. As it
was mentioned already, the pointlike objects are really destroyed
in a Big-Rip singularity only. However, at SFS the infinite tidal
forces appear, and one may worry about the fate of strings
approaching these singularities. It was shown \cite{strings} that this is subtle
in the sense that strings are not infinitely stretched (remain finite
invariant size) at any of these singularities apart from a Big-Rip.
In other words, extended objects like strings, despite
infinite tidal forces, may cross through SFS, GSFS, type III, and
type IV singularities.

\section{Generalized energy conditions and exotic singularities.}
\noindent
From the above considerations it is clear that the application of
the standard energy conditions to exotic singularities is not very
useful. Then, one should try to formulate some different energy
conditions which may be helpful in classifying exotic
singularities \cite{statef}. This may be put in the context of the
higher-order characteristics of the expansion (statefinders) which involve
higher-order derivatives of the scale factor such as jerk, snap
etc. \cite{statef,jerk+snap}. For example, one could think of
a hybrid energy condition like $\alpha \varrho > \dot{p}$ with $\alpha=$
const., to prevent an emergence of SFS, or a higher-order dominant
energy condition in the form $\dot{\varrho} > \mid \dot{p} \mid$, whose
violation can be a good signal of GSFS.

\section{Conclusion}\label{dicu}

\noindent
Universe acceleration gave some motivation to study non-standard
cosmological singularities such as Big-Rip, Sudden Future
Singularity, Finite Density singularity and type III, IV
singularities. However, most of these singularities (apart from Big-Rip) are weak
singularities which do not exhibit geodesic incompletness and
allow the evolution of both pointlike objects and strings through
them.

\section*{Acknowledgments}

This work has partially been supported by the
Polish Ministry of Science and Education grant No 1P03B 043 29 (years
2005-07).




\vfill

\end{document}